\documentclass[prx,twocolumn,showpacs,superscriptaddress]{revtex4-2}

\usepackage{graphicx}
\usepackage{setspace}
\usepackage{amsmath}
\usepackage{mathtools}
\usepackage{tabularx}
\usepackage{mathrsfs}
\usepackage{lipsum}
\usepackage{xcolor}
\usepackage[citecolor=blue,linkcolor=blue,colorlinks=true,linkbordercolor=white]{hyperref}
\usepackage{float}

\newcommand{\be}{\begin{equation}}
\newcommand{\ee}{\end{equation}}
\newcommand{\bea}{\begin{eqnarray}}
\newcommand{\eea}{\end{eqnarray}}

\newcommand{\la}{\langle}
\newcommand{\ra}{\rangle}
\newcommand{\ua}{\uparrow}
\newcommand{\da}{\downarrow}

\begin{document}


\title{Linear entropy fails to predict entanglement behavior in low-density fermionic systems}

\author{T. Pauletti}


\author{M. Garcia}

\author{G. A. Canella}

\author{V. V. Fran\c{c}a}

\affiliation{Institute of Chemistry, S\~{a}o Paulo State University, 14800-090, Araraquara, S\~{a}o Paulo, Brazil}

\begin{abstract}

Entanglement is considered a fundamental ingredient for quantum technologies and condensed matter systems are among the good candidates for quantum devices. For bipartite pure states the von Neumann entropy is a proper measure of entanglement, while the linear entropy, associated to the mixedness of the reduced density matrices, is a simpler quantity to be obtained and is considered to be qualitatively equivalent to the von Neumann. 
Here we investigate both linear and von Neumann entropies for quantifying entanglement in homogeneous, superlattice and disordered Hubbard chains. We find regimes of parameters for which the linear entropy fails in reproducing the qualitative behavior of the von Neumann entropy. This then may lead to incorrect predictions {\it i)} of maximum and minimum entanglement states and {\it ii)} of quantum phase transitions.

\end{abstract}


\maketitle

\section{Introduction}

Entanglement $-$ one of the most remarkable features of quantum mechanics $-$ is conceived as a fundamental resource for quantum information process, including computation, communication and metrology, thus essential to quantum technologies. Entanglement has been studied by several research communities, such as ultracold gases, quantum optics and semiconductors \cite{rev}. Condensed matter systems, in particular, are considered good candidates for quantum technology devices due to their natural quantum fluctuations. 

From the theoretical point of view, entanglement has been widely investigated in the Hubbard model \cite{larsson, zanardi, gu, sarandy, picoli, universal, iemini,physA,v3,v7,v11,v10}, which is one of the most used simplified models to describe strongly interacting fermions \cite{vivaldo}. Concerning entanglement measures, among the general class of Renyi entropies \cite{ref1.30, ref1.31}, the von Neumann \cite{vN} is the most commonly used entropy for quantifying entanglement of bipartite pure states. The von Neumann entropy requires however the diagonalization of reduced density matrices, a numerically demanding task for most of the complex many-body systems. 

Thus entropy measures which are simpler to be calculated, as the linear entropy \cite{ref1.30, ref1.31, ref2.27, ref2.27.5, ref20} $-$ associated to the purity of the reduced system $-$ are extremely valuable from the computational point of view. Another advantage of the linear entropy is to allow analytical approaches \cite{pra11, ref12}. Although the linear entropy is not additive \cite{ref2.27}, in general it is considered qualitatively equivalent to the von Neumann entropy, and therefore has being widely exploited to quantify entanglement and non-classical correlations in several contexts \cite{ref1, ref2, ref2.15, marcos1, v1,v2,v4,v5, marcos2, ref5, ref5.29, ref6, ref7, ref7.5, ref7.6, ref7.7, ref9, ref10, ref11, ref12, ref17, ref18}. 

In this work we compare the two entanglement measures, linear and von Neumann entropies, for quantifying single-site entanglement in homogeneous, superlattice and disordered Hubbard chains. We find regimes of filling factor for which the linear entropy fails miserably to predict the qualitative behavior of entanglement with interaction. This could lead to incorrect predictions of maximum and minimum entanglement regimes and consequently to induce incorrect predictions and/or interpretations of quantum phase transitions.  
\vspace{-0.5cm}
\section{Theoretical Model and Computational Methods}

Here the one-dimensional nanostructures of interacting fermions are described by the single-band Hubbard model, 
\be{}
H = -t\sum_{\la ij \ra \sigma}(\hat{c}^{\dagger}_{i\sigma}\hat{c}_{j\sigma}+ h.c.)  + 
U\sum_i\hat{n}_{i\ua}\hat{n}_{i\da} + \sum_{i\sigma}V_i\hat{n}_{i\sigma},
\ee{}

\noindent where $t$ is the hopping parameter between neighboring sites $<ij>$, $U$ is the on-site Coulomb interaction and $V_i$ is the external potential, which will be used to simulate superlattice and disorder potentials. We adopt $U$ and $V_i$ in units of $t$ and set $t=1$. Here $\hat{c}^{\dagger}_{i\sigma}$ ($\hat{c}_{i\sigma}$) is the creation (annihilation) fermion operator with $z$-spin component $\sigma = \ua,\da$ at site $i$, $\hat{n}_{i\sigma} = \hat{c}^{\dagger}_{i\sigma}\hat{c}_{i\sigma}$ is the particle density operator, $n = N/L$ is the average density or filling factor, $N = N_{\ua} + N_{\da}$ is the total number of particles and $L$ is the chain size. Our chains are restricted to zero temperature, open boundary conditions and fixed number of particles with spin-balanced population, $N_{\ua} = N_{\da}$.

We consider the ground-state single-site entanglement defined as the entanglement between a single site and the remaining $L-1$ sites \cite{zanardi}. For such bipartite entanglement of total pure state, the von Neumann entropy \cite{vN} is established as a proper entanglement measure, 
\begin{eqnarray}
 \mathcal S=-\frac{1}{\ln(d)}Tr[ \rho_i \ln \rho_i]=-\frac{1}{\ln(d)}\sum_{k} \text w_{ik}\ln \text w_{ik},\label{vn}
\end{eqnarray}
\noindent where $\rho_i=Tr_B [\rho]$ is the reduced density matrix of site $i$, obtained via the partial trace over the subsystem $B$ ($L-1$ sites) from the total density matrix $\rho$. In the occupation's basis, $S$ can be Schmidt decomposed in terms of the occupation probabilities $\text w_{ik}$, in which $k=\uparrow, \downarrow, 2,0$ runs over the Hilbert space with dimension $d=4$: $\text w_{i\uparrow}=\text w_{i\downarrow}=n_i/2-\text w_{i2}$ for the single occupation, with spin up or down, $\text w_{i2}= \langle \hat n_{i\uparrow}\hat n_{i\downarrow}\rangle=\partial e_0/\partial U$ for the double occupation, where $e_0(n_i,U)$ is the per-site ground-state energy, and $\text w_{i0}=1-\text w_{i\uparrow}-\text w_{i\downarrow}-\text w_2$ for the empty occupation probability, such that $\sum_k \text w_{ik}=1$.  The term $1/\ln (d)$ is the normalization factor.

\begin{figure}[!t]
\centering
\includegraphics[scale=0.25]{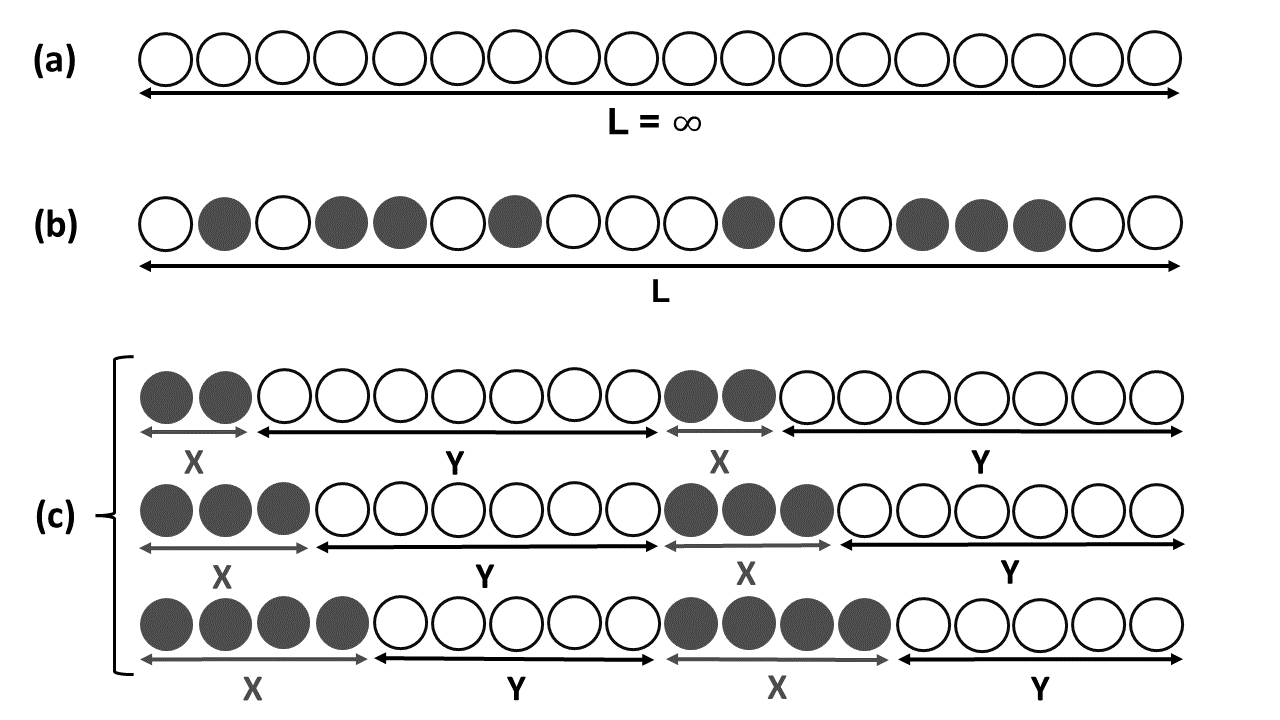}
\caption{Schematic representation of (a) an infinite homogeneous system with $V=0$ (empty circles), (b) a disordered chain, with impurities $V$ randomly distributed (filled circles), and (c) distinct $X:Y$ superlattice structures, with $X$ sites with impurities $V$ and $Y$ sites without impurities. }
\end{figure}

An alternative measure for the entanglement is the linear entropy,
\begin{eqnarray}
\mathcal L =\frac{d}{d-1}\left(1-Tr[\rho_i^2]\right)
=\frac{d}{d-1}\left(1-\sum_{k} \text w_{ik}^2\right),\label{lin}
\end{eqnarray}
\noindent where $d/(d-1)$ is the normalization factor and $Tr[\rho_i^2]$ is the purity of the reduced system. Thus the linear entropy quantifies the mixedness of the reduced density matrices, which for a pure state can arise only from the entanglement between the subsystems. This then justifies a common belief in the quantum information community that $ \mathcal S$ and $\mathcal L$ are qualitatively equivalent. 

Mathematically, by Taylor expanding the von Neumann entropy,
\begin{eqnarray}
 \mathcal S&=&-\frac{1}{\ln(d)}\sum_k \text w_{ik}\left[\sum_l\frac{(-1)^{l+1}}{l}\left(\text w_{ik}-1\right)^l\right],
\end{eqnarray}
one finds that the first order expansion $ \mathcal S_{l=1}\equiv \mathcal S_1$ is proportional to the linear entropy:
\begin{eqnarray}
 \mathcal S_1=\frac{1}{\ln(d)}\left(1-\sum_k \text w_{ik}^2\right)=\frac{d-1}{d\hspace{0.02cm}\ln(d)}\mathcal L.
\end{eqnarray}
Our aim is then to study the qualitative equivalence between $\mathcal L$ and $ \mathcal S$ in three distinct systems, as illustrated in Figure 1: {\it i)} homogeneous chains,  {\it ii)} disordered systems and {\it iii)} superlattices structures.

\begin{center}
\it i) Homogeneous Chains
\end{center}

For the infinite homogeneous chains (Fig. 1a) in which all sites are equivalent, $n_i=n$, we calculate the probabilities $\text w_k$ using the analytical FVC approximation \cite{fvc} for the homogeneous per-site ground-state energy $e_0^{hom}(n,U)\approx e_0^{FVC}(n,U)$,
\begin{eqnarray}
e_0^{FVC}(n,U)=
-\frac{2 \beta(n,U)}{\pi} \sin \left(\frac{\pi n}{\beta(n,U)}\right),
\label{fvc}
\end{eqnarray}
where $\beta(n,U)=b(U)^{\alpha(n,U)}$, $\alpha(n,U)=n^{\sqrt[3]{U}/8}$ and $b(U)$ is determined from
\begin{equation}
-\frac{2b(U)}{\pi}\sin\left(\frac{\pi}{b(U)}\right)=-4\int_0^\infty dx \frac{J_0(x)J_1(x)}{x\left(1+e^{Ux/2}\right)}.
\end{equation}
The $e_0^{FVC}$ expression is valid for $U\geq 0$ and $n\leq 1$, but it can be extended to $n>1$ and to $U<0$ by standard particle-hole transformations \cite{lw1,lw2, schlottmann}. By construction, Eq. (\ref{fvc}) becomes exact for $U\to 0$ and $U\to \infty$ (any $n$), and for $n=1$ (any $U$), and provides a reasonable approximation to the full Bethe-Ansatz solution \cite{lw1,lw2}  inbetween. 

Despite these exact limits of $e_0^{FVC}$, the entanglement measures involve the energy derivative via the double occupancy, $\text{w}_2=\partial e_0^{FVC}/\partial U$, thus $ \mathcal S$ and $\mathcal L$ obtained with FVC are approximations (except at $n=1$, where the derivative is also exact). Besides, we adopt $U=0.2t$ as $U\sim0$, due to high fluctuations in the FVC derivatives for $U\rightarrow 0$. Thus for comparison, in several regimes of parameters we plot also the entanglement obtained via density-matrix renormalization group (DMRG) \cite{dmrg} calculations for large ($L=120$) chains, which is considered almost exact.

Notice that for inhomogeneous chains, with a certain density profile $\{n_i\}$, the entanglement can not be directly obtained through the homogeneous $e_0^{FVC}(n,U)$ expression. Thus for disordered chains and superlattices we adopt different approaches, as detailed below.

\begin{center}
\it ii) Disordered chains and Superlattices
\end{center}

We simulate the disordered chains with a given concentration $C$ of pointlike impurities of same intensity $V$ randomly distributed along the chain (Fig.1b). Entanglement is then calculated via a local density approximation (LDA) for both entropies, as first proposed in Ref. \cite{v11},

\begin{equation}
{ \mathcal S}^{inh} \approx { \mathcal S}^{LDA} \equiv \frac{1}{L}\sum_i{ \mathcal S}(n,U)|_{n \rightarrow n_i},
\end{equation}
\vspace{-0.5cm}%
\begin{equation}
\mathcal{L}^{inh} \approx \mathcal{L}^{LDA} \equiv \frac{1}{L}\sum_i\mathcal{L}(n,U)|_{n \rightarrow n_i},
\end{equation}
\noindent where $ \mathcal S(n,U)$ and $\mathcal{L}(n,U)$ are the homogeneous density functionals given by Eq.(\ref{vn}) and Eq.(\ref{lin}).

\begin{figure}[!t]
\centering
\includegraphics[scale=0.28]{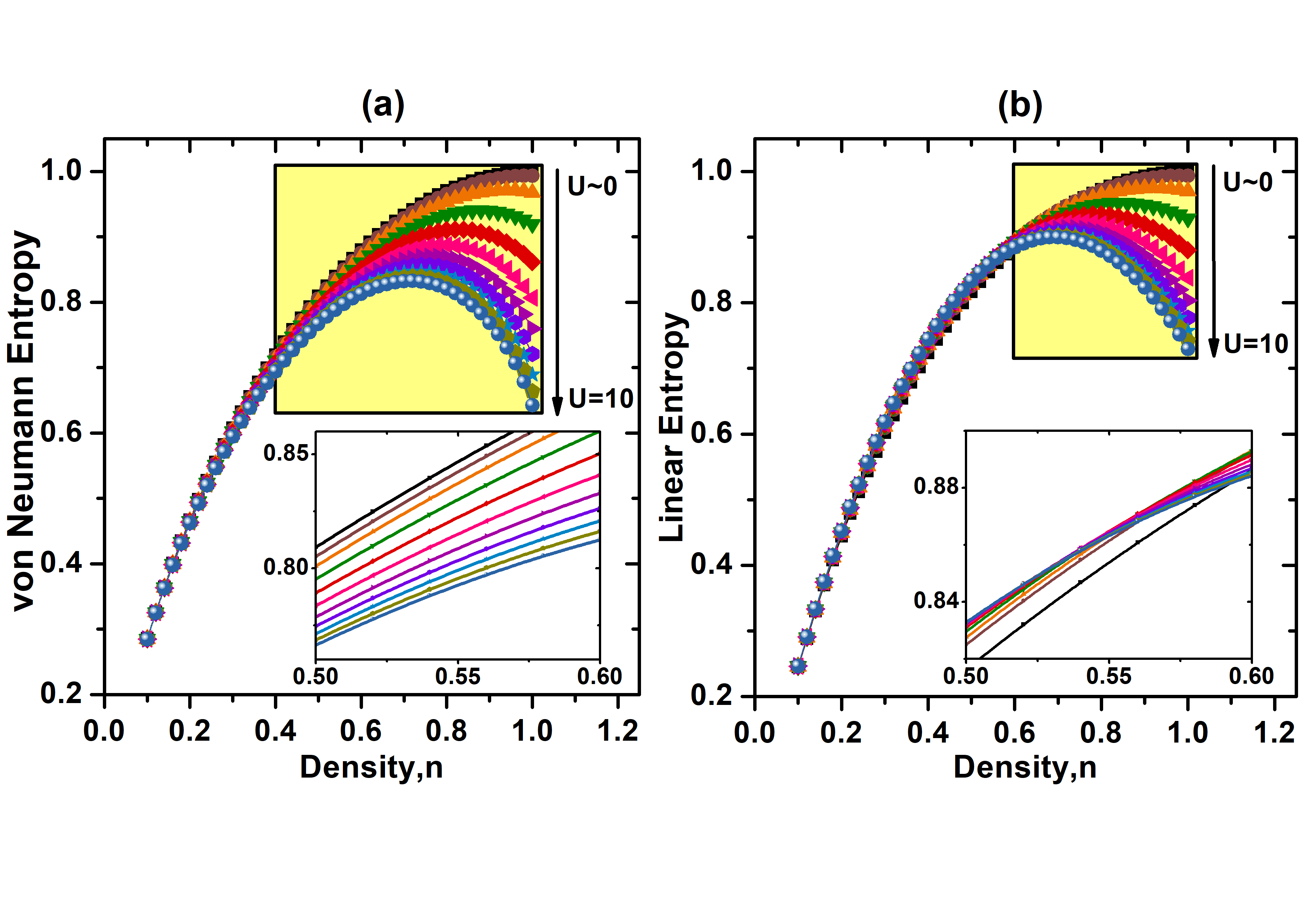}
\vspace{-0.9cm}
\caption{Entanglement of homogeneous infinite Hubbard chains quantified by the von Neumann entropy (a) and the linear entropy (b) as a function of density for several repulsive interactions $U$. The linear entropy is less sensitive to the interaction for $n\leq 0.6$ than the von Neumann entropy. The inset shows that while the von Neumann entropy decreases monotonically with the interaction, the linear entropy in this regime of low density, gives the incorrect trend of a increasing entanglement with $U>0$.}
\end{figure}

To ensure that the results for disordered chains are not dependent on specific configurations of impurities, we generate $100$ samples for each set ($U,n; C,V$) of parameters. Thus the entanglement considered in this case is not only averaged over the sites ($L=100$), but also averaged over the 100 samples. This huge amount of data makes it impracticable to use exact methods such as DMRG to obtain the probabilities $w_{ik}$. So we use instead standard Kohn-Sham Density Functional Theory calculations for the Hubbard model \cite{vivaldo} (within LDA for the exchange-correlation energy, using Bethe-Ansatz numerical solution as input) to obtain the density profile $\{n_i\}$. We also adopt a LDA approach for the double occupancy, using the FVC functional as input,

\begin{equation}
w_{i2}\approx w_{i2}^{LDA}=\left.\frac{\partial e_0^{FVC}(n,U)}{\partial U}\right|_{n\rightarrow n_i}.
\end{equation}

For the superlattice structures \cite{ref21,sl, sl2} our results are obtained from exact DMRG calculations for finite chains ($L=36$) with a periodical external potential: $V_i=V$ for the first $X$ sites and $V=0$ for the next $Y$ sites, thus composing the superlattice modulation $\text X:Y$ (Fig. 1c). 

\section{Results and Discussion}

In Figure 2 we present a first comparison between $ \mathcal S$ and $\mathcal L$ as a function of $n$ for the infinite homogeneous Hubbard chain. First, we see that both entropies have the expected non-monotonic behavior with $n$.  This non-monotonicity when approaching the critical density $n=1$ is precisely how entanglement signs the Mott metal-insulator transition \cite{larsson,v0}: entanglement increases with the number of particles in the metallic phase, while in the Mott-localized state entanglement decreases due to the suppression of doubly and empty states. Second, we find that for low densities the curves for distinct $U$'s lie essentially on top of each other, becoming distinguishable only for higher $n$. But clearly the linear entropy is less sensitive to the Mott MIT than the von Neumann entropy: while $S$  distinguishes between $U$'s for $n\gtrsim 0.4$, $\mathcal L$ does it only for $n\gtrsim 0.6$, as highlighted in yellow.  

Another well known feature of the single-site entanglement in the Hubbard model is the fact that the maximum entanglement occurs for the non-interacting case, $U=0$, and that by increasing $U$ entanglement decreases monotonically \cite{larsson,v0}. This reflects the fact that at $U=0$ the four occupation probabilities are in their best balance: no particular state is preferred and therefore the single-site entanglement is maximum for a fixed density $n$.  By increasing the repulsive $U$, the unpaired probabilities $\text w_\uparrow$, $\text w_\downarrow$ are favored, what then leads to lower entanglement. 

\begin{figure}[!t]
\centering
\includegraphics[width=9cm]{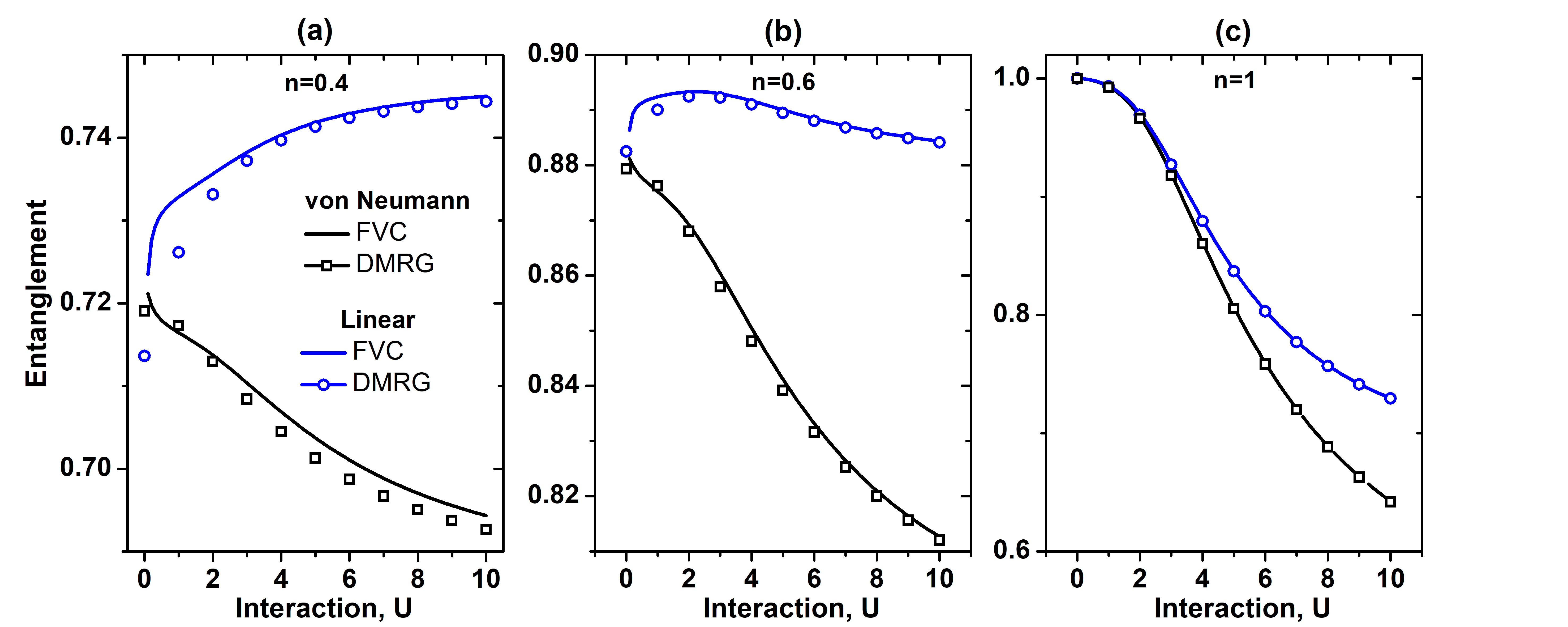}
\caption{Entanglement of homogeneous infinite Hubbard chains quantified by both von Neumann and linear entropies using FVC approximation as a function of interaction for low density (a), intermediate density (b) and at half filling (c). In all panels filled symbols represent DMRG calculations for finite chains with $L=120$ sites.}
\end{figure}

However, Figure 3 (see also the insets of Fig. 2) reveals that while $\mathcal S$ properly captures this physics for any $n$, the linear entropy fails to recover this monotonic decreasing of entanglement with $U$ for $n\lesssim 0.6$.  For low densities (Fig. 3a), $\mathcal L$ provides precisely the opposite of the expected behavior: minimum entanglement at $U=0$ increasing monotonically as $U$ increases. For intermediate filling factors  (Fig. 3b), $\mathcal L$ incorrectly predicts a non-monotonic behavior with $U$. The qualitative behavior of $\mathcal S$ is only recovered by $\mathcal L$ for $n>0.6$ (Fig. 3c), becoming worse for strong interactions ($U\gtrsim 5t$). To guarantee that the failure of the linear entropy has no relation to the FVC parametrization, we also present in the three panels of Fig. 3 DMRG data for a finite but large ($L=120$) chain: the overall agreement then discard a possible artifact of the FVC approximation.

\begin{figure}[!t]
\centering
\hspace{-0.2cm}\includegraphics[scale=0.3]{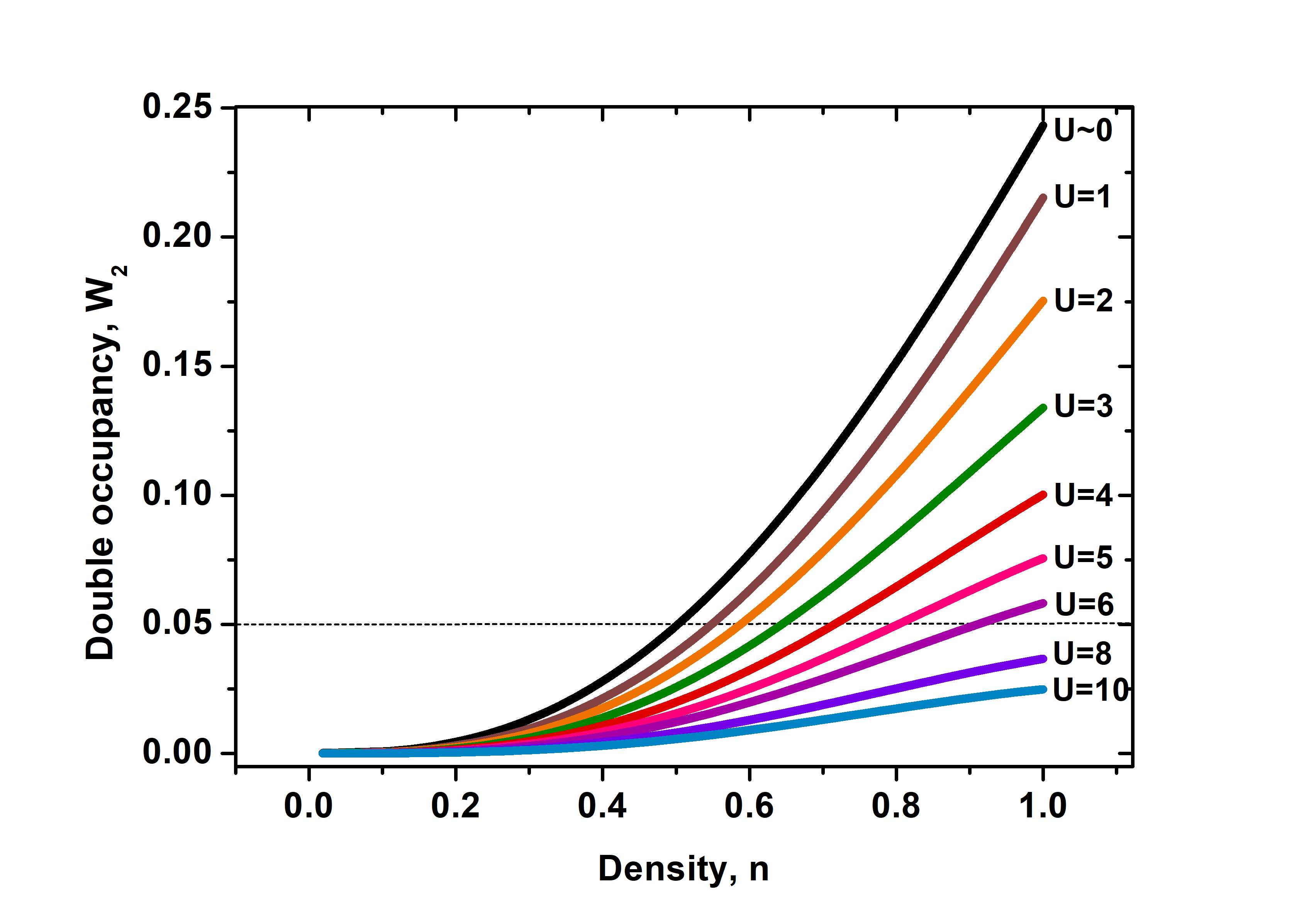}
\caption{Double occupation probability $\text w_2\equiv \partial e_0/\partial U$ as a function of the density for several interactions as obtained by the FVC approximation.}
\end{figure}

In order to understand the poor performance of the linear entropy for $n\lesssim 0.6$, we analyze in Figure 4 the double occupation probability $\text w_2$ as a function of $n$ for several $U$. We see that the relevance of $\text w_2$ depends on both $n$ and $U$. For $n\lesssim 0.6$, $\text w_2$ is almost independent on the interaction and essentially negligible, as the chain is sufficiently empty for accommodating unpaired particles, thus avoiding the additional energetic cost due to $U>0$.  The probabilities in this density regime are then: $\text w_2\sim 0$, $\text w_\uparrow=\text w_\downarrow\sim n/2$ and $\text w_0\sim1-n$ and, therefore, the logarithm expansion (Eq.(4)) will contain terms proportional to $(-n)^k$ and $(n/2-1)^k$. So for the terms $(-n)^k$, as $n\leq 0.6$ one can clearly discard the contributions coming from higher orders ($k>1$). In contrast, for $(n/2-1)^k$ terms, since the base ranges from $0.7$ to $1.0$ it is essential to expand beyond the linear term. This then explains why $\mathcal L$ fails miserably to recover $\mathcal S$ {\it even qualitatively} for $n\lesssim 0.6$. 

Exactly the same argument holds also for strong $U\gtrsim 5t$ and higher densities $n\gtrsim 0.6$, since $\text w_2\rightarrow 0$. This interpretation is confirmed also in the regime of $n\gtrsim 0.6$, where $\mathcal L$ properly captures the physics: Fig. 3c shows that while $\mathcal L$ is {\it quantitatively} equivalent to $\mathcal S$ for $U\lesssim 5t$, it performs worse for $U\gtrsim 5t$. 

Figure 5 reveals that for $n=0.5$ one needs to consider up to the sixth order ($\mathcal S_6$) in the expansion for recovering qualitatively the entanglement, while for $n=0.2$ the $\mathcal S_{25}$ is required. Our results suggests that although all the mixedness of the $L-1$ sites, quantified via $\mathcal L$, is necessarily due to entanglement, there are more information about the correlations $U$ in the entanglement between the subsystems than in the mixedness. From the model point of view, this tell us that the energy dependence on the condensed matter correlations $U$ in this density $n\lesssim 0.6$ regime is so small, that discarding higher order terms impacts significantly the entanglement measure.



\begin{figure}[!t]
\centering
\hspace{0cm}\includegraphics[scale=0.21]{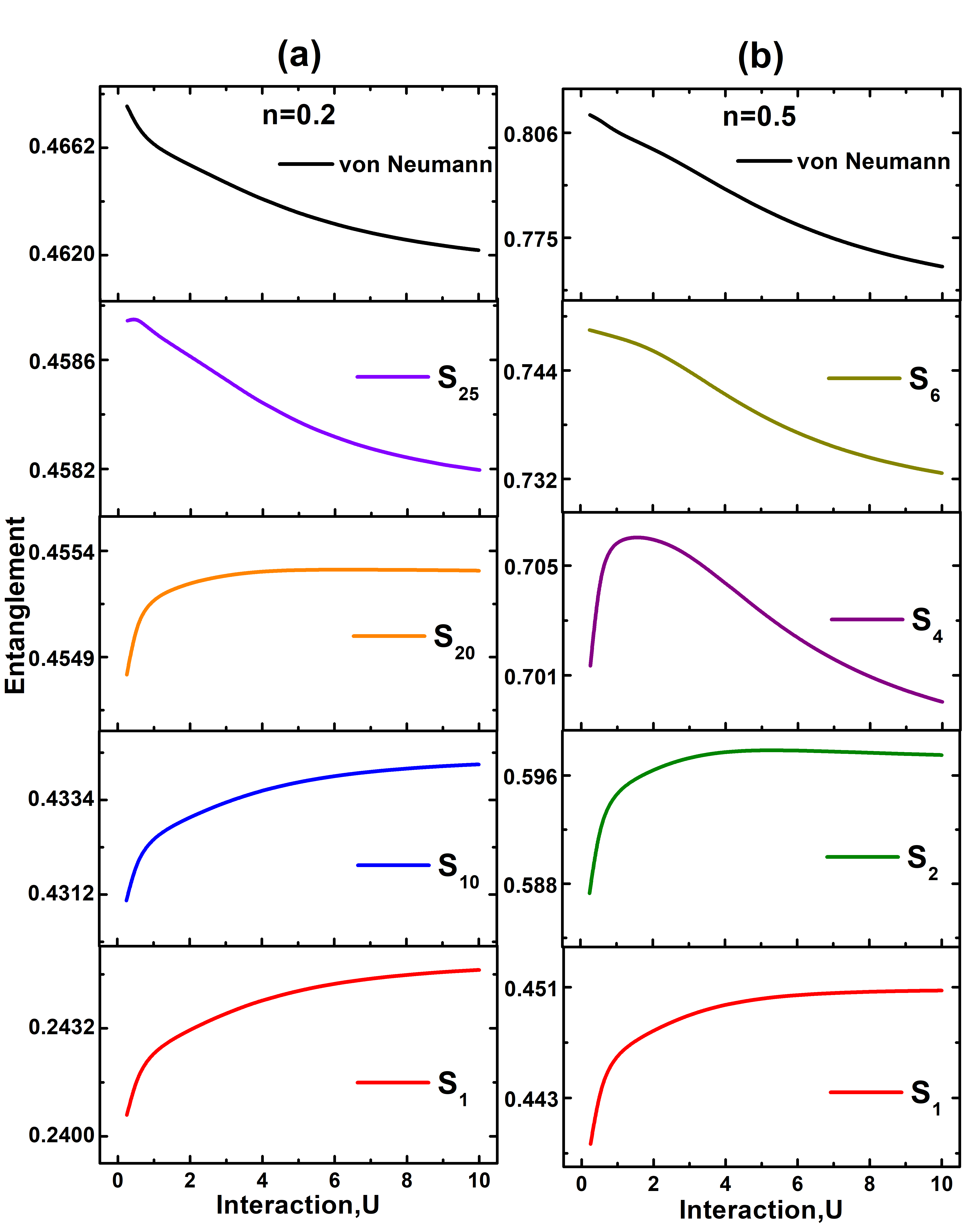}
\caption{Analysis of the $l-order$ Taylor expansion of the von Neumann entropy $\mathcal S_l$ for homogeneous infinite Hubbard chains as a function of $U$ at (a) $n=0.5$  and (b) $n=0.2$.}
\end{figure}

Up to this point, by studying homogeneous Hubbard chains, it is clear that there are specific regimes of $n$ and $U$ for which the linear entropy fails to recover the behavior of the entanglement. Now inhomogeneous systems, such as disordered chains and superlattices, imply in a highly inhomogeneous density profile $\{n_i\}$. Within the LDA for entropy concept (Eqs. (8) and (9)), the average single-site entanglement will then be quantified through an average of entropies for these different densities $n_i$, which in turn may range from regimes for which $\mathcal L$ fails to  regimes for which $\mathcal L$ succeeds. Thus, concerning the limitation of the linear entropy for low density regimes, our question is: can one trust on the linear entropy to describe entanglement in inhomogeneous systems?

\begin{figure}[!t]
\centering
\includegraphics[scale=0.9]{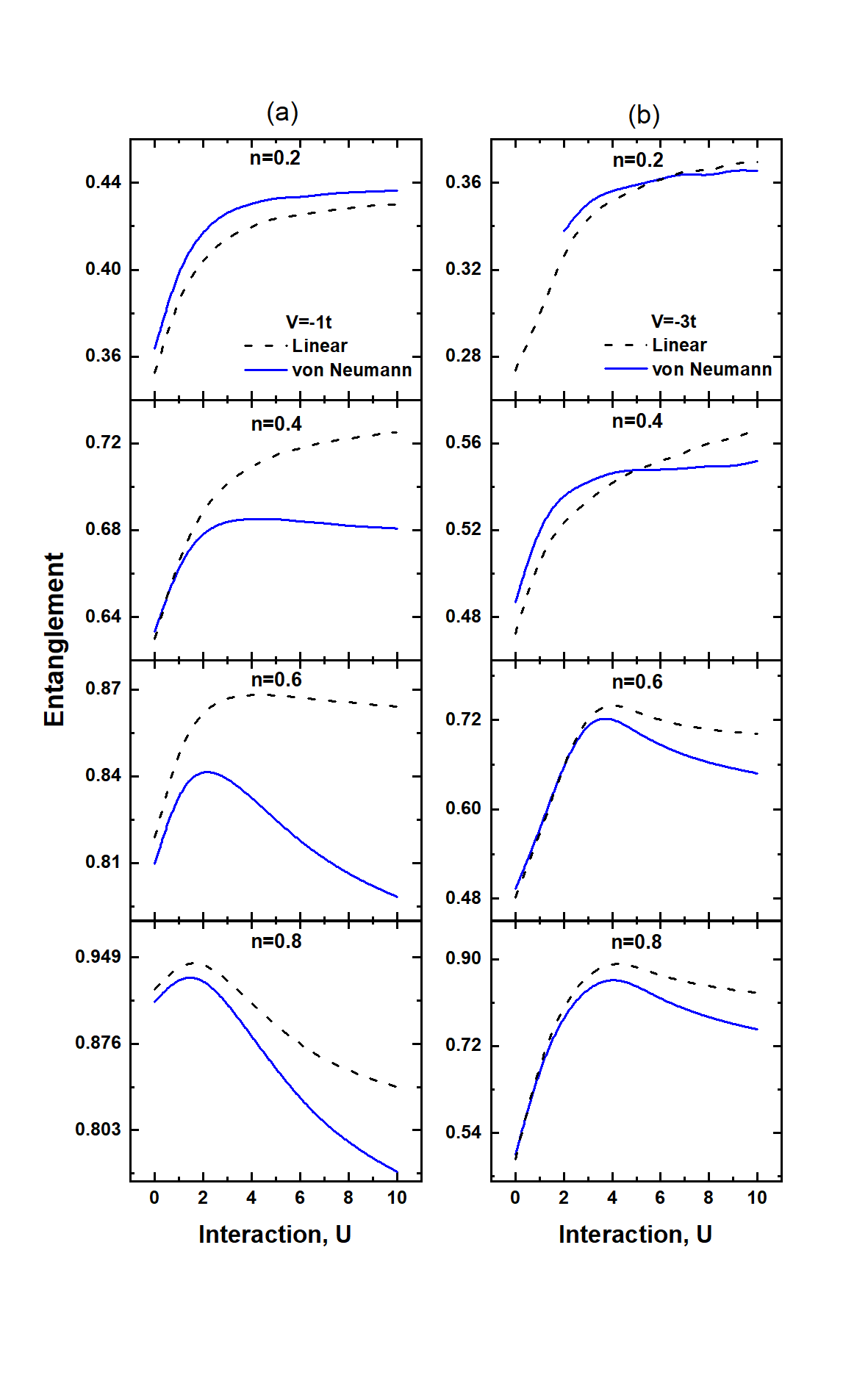}
\vspace{-1.5cm}
\caption{Entanglement of finite disordered chains quantified by the von Neumann and linear entropies as a function of the interaction for a concentration of $C=40\%$ impurities with disorder strength (a) $V=-1t$ and (b) $V=-3t$.}
\end{figure}

We start this analysis by considering disordered chains composed of randomly distributed point-like impurities $V$ (illustrated in Fig. 1b). Figure 6 compares $S$ and $\mathcal L$ as a function of $U$ for several $n$ and two disorder intensities. At $U=0$ for any density and $V$ the system is Anderson localized and therefore has a considerably small entanglement (compared to the clean $V=0$ case). By turning the interaction on the entanglement increases due to the competition between $V$ and $U$. Fig. 6a reveals that this entanglement increasing is a small effect for low densities ($n\lesssim0.4$) and quickly saturates, since at this regime the interaction plays a less relevant role. Both entropies recover properly this scenario. 

Nevertheless, for stronger densities, Fig. 6a shows that the competition between $U$ and $V$ leads to a higher change in the entanglement degree: it reaches a maximum and thus decreases saturating at a finite value. This effect in this case is stronger for larger $n$, since then the interaction becomes more relevant. Our results show however that the linear entropy is less sensitive to this feature: at $n=0.4$ $\mathcal L$ still does not capture the non-monotonic behavior of $S$, while at $n=0.6$ $\mathcal L$ reproduces the non-monotonic trend but in a smoother way. As the density increases further ($n\gtrsim 0.6$) we find that $\mathcal L$ recovers better the $S$ behavior, similarly to what we observed in the homogeneous case. 

Interestingly by increasing the disorder intensity $V$, Fig. 6b, $\mathcal L$ is able to recover the qualitative behavior of $S$ for all the density regimes. We attribute this very good performance of the linear entropy for stronger disorder $V$ to the fact that within this regime the sites are either strongly populated, with high $n$ in the impurity sites (for $V<0$) or empty in the non-impurity sites. Thus, even for the low average density regime, $n=0.4$, in which the homogeneous $\mathcal L$ would be very poor, for the disordered chain there are effectively stronger densities $n>0.4$ and thus $\mathcal L$ performs better.  We have also analyzed the entropies for distinct concentrations (not shown) but there was no significative change on the trends discussed above.

\begin{figure}[!t]
\centering
\includegraphics[scale=0.17]{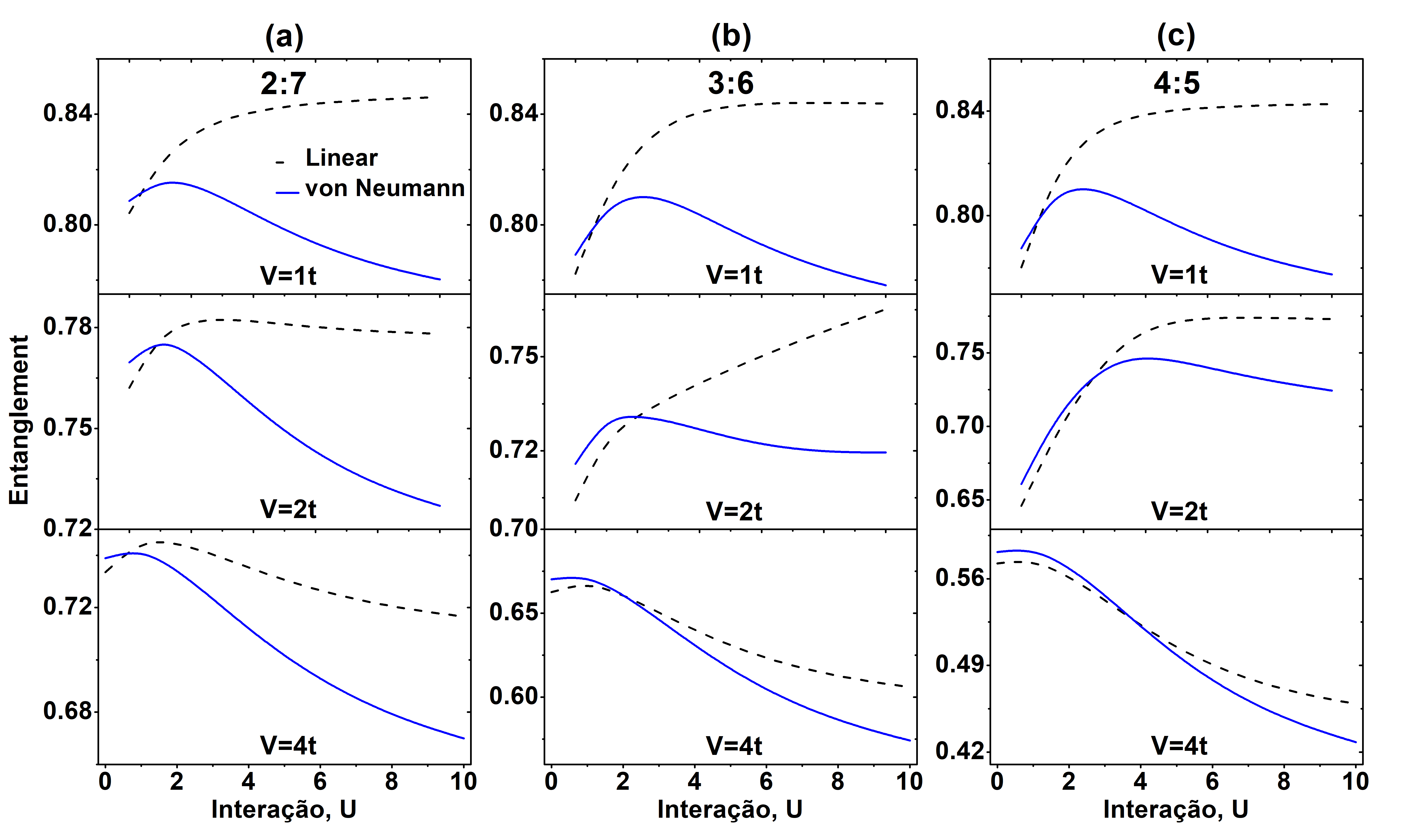}
\caption{Entanglement of finite superlattices quantified by the von Neumann and linear entropies as a function of the interaction  for several superlattice potentials $V$ for distinct superlattice structures 2:7 (a), 3:6 (b) and 4:5 (c). In all cases the unitary cell has size $9$, the chain has $L=36$ sites, with fixed number of particles $N_\uparrow=N_\downarrow=10$. }
\end{figure}

Finally in Figure 7 we analyze the performance of $\mathcal L$ in superlattices (illustrated in Fig. 1c). We find that the linear entropy fails to describe the non-monotonic behavior of entanglement with interaction for small potential strengths ($V=1t$, $V=2t$) for all the superlattice modulations $X:Y$. As the superlattice potential $V$ increases and/or the balance between impurity and non impurity sites increases (for $X\rightarrow Y$) the linear and the von Neumann entropies become qualitatively alike. Similarly to the disordered case, this reflects the fact that with a stronger superlattice potential the density profile $\{ni\}$ is farther from the average density n, thus lying in a regime where $\mathcal L$ properly recovers the entanglement.

\section{Conclusion}

Our results comparing the linear entropy to the von Neumann entropy in the Hubbard model clearly prove that the common belief that both entropies are qualitatively equivalent is not always true. We find that for small densities ($n\lesssim 0.6$), the linear entropy fails to predict the qualitative behavior of entanglement with the interaction. This is particular relevant for many-body interacting (and inhomogeneous) systems, where by simplicity one often chooses to use the linear entropy as an entanglement quantifier: it may lead to {\it i)} incorrect predictions of maximum and minimum entanglement, what could then induce to {\it ii)} incorrect predictions of quantum phase transitions. 

From a more fundamental point of view, the fact that the 1D Hubbard model is substantially sensitive to the two entropies suggests one could use the model as a test bed for studying differences between other entanglement measures and other quantum information properties, what certainly could benefit the development of quantum technologies.

\begin{center}\bf ACKNOWLEDGMENTS \end{center}

We thank enlightening discussions with L. Celeri, M. C. Oliveira and C. Villas-Boas. This research was supported by FAPESP (2021/06744-8), CNPq (403890/2021-7; 140854/2021-5), Coordena\c{c}\~{a}o de Aperfei\c{c}oamento de Pessoal de Nivel Superior - Brasil (CAPES) -
Finance Code 001, and by resources supplied by the Center for Scientific Computing (NCC/GridUNESP) from S\~{a}o Paulo State University. 



\begin{thebibliography}{10}
\bibitem{rev}N. Friis, G. Vitagliano, M. Malik, and M. Huber, Nature Reviews Physics {\bf1}, 72 (2019).
\bibitem{zanardi} P. Zanardi, Phys. Rev. A {\bf65}, 042101(R) (2002).
\bibitem{v11} V. V. Fran\c{c}a and K. Capelle, Phys. Rev. Lett. {\bf100}, 070403 (2008).
\bibitem{larsson} D. Larsson and H. Johannesson, Phys. Rev. Lett. {\bf95},
196406 (2005).
\bibitem{v3}D. Arisa and V. V. Fran\c{c}a, Phys. Rev. B {\bf101}, 214522 (2020).
\bibitem{v7} J. P. Coe, V. V. Fran\c{c}a, and I. D'Amico, EPL {\bf93}, 10001 (2011).	
\bibitem{gu} S.-J. Gu, S.-S. Deng, Y.-Q. Li, and H.-Q. Lin, Phys. Rev. Lett. {\bf93}, 086402 (2004).
\bibitem{sarandy} L.-A. Wu, M. S. Sarandy, and D. A. Lidar, Phys. Rev. Lett.
{\bf93}, 250404 (2004).
\bibitem{universal}  V. V. Fran\c{c}a and K. Capelle, Phys. Rev. A {\bf77}, 062324 (2008).
\bibitem{iemini} D. L. B. Ferreira, T. O. Maciel, R. O. Vianna, and F. Iemini,
Phys. Rev. B {\bf105}, 115145 (2022).
\bibitem{physA} V. V. Fran\c{c}a, Phys. A {\bf475}, 82 (2017).
\bibitem{picoli} T. de Picoli, I. D'Amico, and V. V. Fran\c{c}a, Braz. J. of Phys. {\bf48}, 472 (2018).
\bibitem{v10} J. P. Coe, V. V. Fran\c{c}a, and I. D'Amico, Phys. Rev. A {\bf81}, 052321 (2010).
\bibitem{vivaldo} K. Capelle, V. L. Campo Jr, Phys. Rep. {\bf528}, 91 (2013).
\bibitem{ref1.30} A. Renyi, in Proceedings of the Fourth Berkeley Symposium on
Mathematics, Statistics and Probability, Vol. I (University of
California, Berkeley, 1961), p. 547.
\bibitem{ref1.31} J. C. Principe, Information Theoretic Learning: Renyis Entropy
and Kernel Perspectives, Information Science and Statistics
Series (Springer, New York, 2010).
\bibitem{vN} J. von Neumann, The Mathematical Foundations of Quantum
Mechanics (Princeton University Press, Princeton, NJ, 1955).
\bibitem{ref2.27.5} W.H. Zurek, S. Habib, and J.P. Paz, Phys. Rev. Lett. {\bf70}, 1187
(1993).
\bibitem{ref2.27} G. Manfredi and M. R. Feix, Phys. Rev. E {\bf62}, 4665 (2000).
\bibitem{ref20}S. Morelli, C. Klockl, C. Eltschka, J. Siewert, M. Huber,
Linear Algebra and its Applications {\bf584}, 294
(2020).
\bibitem{pra11}V. V. Fran\c{c}a and I. D\' Amico, Phys. Rev. A {\bf83}, 042311 (2011).
\bibitem{ref12} Y. Maleki and A. M. Zheltikov, Opt. Express {\bf27}, 8291-8307 (2019).
\bibitem{ref1}L. E. C. Rosales-Zarate and P. D. Drummond, Phys. Rev. A  {\bf84}, 042114 (2011).
\bibitem{ref2} F. Buscemi, P. Bordone, and A. Bertoni
Phys. Rev. A {\bf75}, 032301(2007).
\bibitem{ref2.15} F. Buscemi, P. Bordone, and A. Bertoni, Phys. Rev. A {\bf73},
052312 (2006).
\bibitem{marcos1} G. Rigolin, T. R. de Oliveira, and M. C. de Oliveira, Phys. Rev. A {\bf74}, 022314 (2006).
\bibitem{marcos2} T. R. de Oliveira, G. Rigolin, and M. C. de Oliveira, Phys. Rev. A {\bf73}, 010305(R) (2006).
\bibitem{v2}G. A. Canella and V. V. Fran\c{c}a, Phys. Rev. B {\bf104}, 134201 (2021).
\bibitem{ref5} F. Benabdallah, A. Slaoui, and M. Daoud, Quantum Inf Process {\bf19}, 252 (2020).
\bibitem{ref5.29} Ma, Z., Chen, Z., Fanchini, F.F., Fei, S.M., Sci. Rep. {\bf5}, 10262 (2015).
\bibitem{ref6} Q. H. Liao, W. J. Nie, J. Xu, Y. Liu, N. R. Zhou, Q. R. Yan, A. Chen, N. H. Liu and M. A. Ahmad, Laser Phys. {\bf26}, 055201 (2016).
\bibitem{v1}G. A. Canella, K. Zawadzki and V. V. Fran\c{c}a, Sci. Rep. {\bf12}, 8709 (2022).
\bibitem{ref7} T. Maciaek and A. Sawicki, J. Phys. A: Math. Theor. {\bf48} 045305 (2015).
\bibitem{ref7.5} A. Olaya-Castro, N. F. Johnson and L. Quiroga, J. Opt. B: Quantum Semiclass. Opt. {\bf6} S730 (2004).
\bibitem{ref7.6}F. Buscemi, P. Bordone and A. Bertoni, Phys. Rev. A {\bf75} 032301 (2007).
\bibitem{ref7.7} C-H. Lin C-H, Y-C. Lin and Y. K. Ho, Few-Body Systems {\bf54}
2147 (2013).
\bibitem{v4}G. A. Canella and V. V. Fran\c{c}a, Phys. A {\bf545}, 123646 (2020).
\bibitem{ref9} Zhi-Yong Ding, H. Yang, H. Yuan, D. Wang, J. Yang, and L. Ye, Phys. Rev. A {\bf101}, 022116 (2020).
\bibitem{ref10} X.-W. Hou, J.-H. Chen, M.-F. Wan, Z.-Q. Ma, Optics Comm. {\bf266} 727 (2006).
\bibitem{ref11} F. Benabdallah and M. Daoud, Eur. Phys. J. D {\bf75}, 3 (2021).
\bibitem{ref17} X. Zha, I. Ahmed, D. Zhang and Y. Zhang,  Laser Phys. {\bf30} 035201 (2020).
\bibitem{ref18} X. Zheng, S.-Q. Ma, G.-F. Zhang, Ann. Phys. (Berlin) {\bf532}, 1900320 (2020).
\bibitem{v5}G. A. Canella and V. V. Fran\c{c}a, Sci. Rep. {\bf9}, 15313 (2019).
\bibitem{fvc} V. V. Fran\c{c}a, D. Vieira, and K. Capelle, New J. of Phys.  {\bf14}, 073021 (2012).
\bibitem{lw1} E. H. Lieb and F. Y. Wu, Phys. Rev. Lett. {\bf 20}, 1445 (1968).
\bibitem{lw2} F. H. Essler et. al, {\it The one-dimensional Hubbard model}, Cambridge University Press (2005).
\bibitem{schlottmann} P. Schlottmann, Int. J. Mod. Phys. B {\bf 11}, 355 (1997).
\bibitem{dmrg} U. Schollw\"ock, Rev. Mod. Phys. {\bf77}, 259 (2005).
\bibitem{ref21} A. D. N. James, M. Aichhorn, J. Laverock, Phys. Rev. Research {\bf3}, 023149 (2021).
\bibitem{sl} T. Br\"unner, E. Runge, A. Buchleitner, V. V. Fran\c{c}a, Phys. Rev. A {\bf87}, 032311 (2013).
\bibitem{sl2}  T. Mendes-Santos, T. Paiva, and R. R. dos Santos, Phys. Rev. B
{\bf87}, 214407 (2013).
\bibitem{v0} V. V. Fran\c{c}a and K. Capelle, Phys. Rev. A {\bf74}, 042325 (2006).
\end{thebibliography}
\end{document}